\documentstyle[aps,prb,graphicx,amsbsy,amssymb]{revtex}

\begin{document}

\wideabs{

\title{Critical Current and Topology of the Supercooled Vortex State in
NbSe$_{2}$}
\author{M. Menghini, Yanina Fasano and F. de la Cruz}
\address{Instituto Balseiro and Centro At\'{o}mico Bariloche, CNEA, Av. Bustillo
9500, Bariloche, RN, Argentina}

\date{August 9, 2001}
\maketitle

\begin{abstract}

We study the behavior of the critical current, $I_{c}(H,T)$, of
pure and Fe doped NbSe$_{2}$ crystals  in the denominated
disordered vortex region, limited by the critical field
$H_{c2}(T)$ and the field $H_{p}(T)$ at which the peak effect in
$I_{c}(H,T)$ is detected. The critical current follows an
individual pinning response as demonstrated by its field
independent universal function of the superfluid density.
Transport measurements combined with Bitter decorations show no
evidence of the existence of an amorphous phase in the high
temperature region.

\end{abstract}


}

\section{Introduction}

Although the peak effect in the critical current, $I_{c}(H,T)$, at
a field $H_{p}(T)$ close to the upper critical field, $H_{c2}(T)$,
has been observed years ago its origin is still subject of debate.
The traditional interpretation of this effect is based on the
explanation suggested by Pippard \cite{pippard} and on the formal
treatment made by Larkin and Ov\-chi\-nni\-kov (LO)\cite{LO}. In
this scenario the peak effect appears as a response of the vortex
lattice to the pre\-sen\-ce of quenched disorder. The peak occurs
due to the softening of the elastic constants in a crossover from
local to non-local elasticity when approaching $H_{c2}(T)$. Using
this formalism the behavior of $I_{c}(H,T)$ at the peak in
NbSe$_{2}$ has been explained.\cite{kes}

The elastic description of the peak effect has been
cri\-ti\-ci\-zed by other authors \cite{batta} emphasizing the
necesary dynamic aspects of the phenomenon. An extensive analysis
through transport measurements in NbSe$_{2}$ showed \cite{batta}
the plastic fingerprint predicted in $2D$
simulations\cite{jensen,kolton} in the $H\!-\!T$ region where the
peak effect appears. In Ref.\onlinecite{batta} a phase diagram
with a liquid vortex phase for $H\!>\!H_{p}(T)$ is proposed.

Recent electrical transport measurements using Corbino-contact
configuration have suggested \cite{corbino} other interpretation
of the peak effect phenomenon. An abrupt decrease of $I_{c}(H,T)$
at $H_{p}(T)$ is considered to be the ma\-ni\-fes\-ta\-tion of a
genuine first order thermodynamic phase transition from a
disordered ($H\!>\!H_{p}(T)$) to an ordered vortex state
($H\!<\!H_{p}(T)$). In this scenario the width of the transition
detected using the conventional four contact configuration is
due\cite{paltielnature} to the injection of a disordered vortex
phase through the surface. It is widely
accepted\cite{banerjee,xiao} that the critical current is history
dependent for \mbox{$H < H_{p}(T)$}  while for
\mbox{$H_{p}(T)\!<\!H\!<\!H_{c2}(T)$} it is reversible.  A
dramatic e\-xam\-ple of history dependence of $I\!-\!V$ curves has
been recently reported \cite{xiao} using a fast transport
measurement technique. This technique allows to define a large
critical current when the sample is cooled, in the presence of a
field (FC), down to temperatures well below that of the peak in
$I_{c}(H,T)$. On the other hand, when u\-sing a slow transport
technique a low critical current is detected. When the field is
a\-pplied after zero field coo\-ling the sample (ZFC) to
temperatures below that of the peak in the critical current, the
same low value of $I_{c}(H,T)$ is obtained independently of the
technique, fast or slow, used to measure it.

Within the frame of this work the most relevant proposed
properties of the mixed state associated with the presence of the
peak effect can be summarized as follows: (a) The vortex phase
nucleates at $H_{c2}(T)$ as a to\-po\-lo\-gi\-ca\-lly disordered
structure with a large critical current that remains in
thermodynamic equilibrium, down to the temperature where the peak
in $I_{c}(H,T)$ takes place; (b) when the sample is FC to lower
temperatures  the dis\-or\-de\-red vortex structure is supercooled
as a metastable state with a large $I_{c}(H,T)$; (c) the
thermodynamic equilibrium state obtained by ZFC experiments at low
tem\-pe\-ra\-tu\-res corresponds to a topologically ordered phase
with low critical current.

Independently of the different interpretations of the peak effect
there is consensus on the existence of an equilibrium disordered
vortex state at temperatures above that of the peak, where
transport measurements show no history dependence. It has been
claimed\cite{corbino,banerjee} that the disordered equilibrium
phase shows a reentrance at fields low enough. In particular, in
Fe doped samples the reentrant region has been found a a field of
$1000\,$Oe.\cite{corbino}

The history independence of the transport properties when
$H\!>\!H_{p}(T)$ suggests that the vortex structure is displaced
homogenously through the sample when the measuring current exceeds
the critical one. However, the analysis of the field and
temperature dependence of $I_{c}(H,T)$ in the so called disordered
phase is still lacking. On the other hand, the claim that in FC
experiments the disordered phase remains in a metastable state at
low temperatures provides us the possibility of using the Bitter
decoration to visualize its topology. This is particularly
important because despite the claim of the presence of a
disordered phase based on the analysis of $I_{c}(H,T)$
measurements, there is no trivial
relation\cite{yaron,duarte,pardo,giamarchi,comment} between
critical current and topological order. The critical current is
determined by vortex displacements in the small length scale
associated with the range of forces of the pinning potential
(superconducting coherence length $\xi(T)$), while the vortex
topology responds to a larger length scale (lattice parameter),
see Ref.\onlinecite{giamarchi}.

In this work we focus the attention on the study of the field and
temperature dependence of the critical current and its relation
with the corresponding vortex structure of the proposed disordered
phase. This is accomplished by complementing the transport
measurements with Bitter decoration of the pristine FC vortex
structure before any external force is applied. Thus, the Bitter
image depicts the vortex structure tested by fast transport
measurements.

\section{Larkin-Ovchinnikov Theoretical Background}

Since the LO theory has been shown \cite{kes} to be
a\-ppro\-pria\-te to provide a quantitative understanding of
$I_{c}(H,T)$ in the peak effect, we take the theoretical
prediction of the model as the basic ingredient for the analysis
of $I_{c}(H,T)$.

This model predicts an enhancement of the critical cu\-rrent when
approaching $H_{c2}(T)$ due to the softening of the vortex lattice
associated with the dispersive cha\-rac\-ter of the tilt modulus.
The increase of $J_{c}$ is a\-sso\-cia\-ted with a monotonous
reduction of the LO correlation volume $V_{c}\!=\!
R_{c}^{2}L_{c}$, where $R_{c}$ is the elastic correlation length
perpendicular to $H$ and $L_{c}$ is the parallel one. At
$H_{p}(T)$, $R_{c}$ takes the minimum value compatible with the
existence of a finite density of vortices, $R_{c}\!=\!a_{0}$,
where $a_{0}\!=\!1.075(\Phi _{0}/B)^{1/2}$ is the vortex lattice
constant. In this picture it is assumed\cite{kes}
$R_{c}\!=\!a_{0}$ in the whole region of fields
$H_{p}(T)\!<\!H\!<\!H_{c2}(T)$ or equivalently
$T_{p}(H)\!<\!T\!<\!T_{c2}(H)$.

In the LO theory the critical force is given by
\mbox{$BJ_{c}V_{c}\!=\!fN^{1/2}$}, where $N$ is the effective
number of pinning centers in $V_{c}$ and the pinning force, $f$,
is usually assumed to be proportional to $\Psi ^{2}$, where $\Psi$
is the Ginzburg-Landau order parameter. Close to $H_{c2}(T)$ the
condensation energy decreases as $\Psi ^{2}\!\propto\!(1-b)(1-t)$
where $t\!=\!T/T_{c}$ and $b\!=\!B/H_{c2}(T)$. It is easily seen
that $(1-b)(1-t)\!=\!\Theta\,T_{c}$ where
$\Theta\!=\!T-T_{c2}(H)$. Within these considerations and using
the non-local elastic cons\-tant
\mbox{$C_{44}^{\ast}\!=\!C_{44}(1-b)(1-t)/\kappa ^{2}$}, where
$C_{44}\! \approx\!H^{2}$  is the local tilting elastic
constant\cite{beqh}, it is found\cite{LO} that for
$H\!>\!H_{p}(T)$

\begin{equation}
V_{c}=L_{c}a_{0}^{2}=a_{0}^{2}(4 C_{44}^{\ast}
a_{0}^{2}/fn^{1/2})^{2/3}\propto H^{-1/3} \label{vc}
\end{equation}

\noindent where it has been assumed that close to $H_{c2}(T)$ all
pi\-nning centers in $V_{c}$ are effective: $N\!=\!nV_{c}$. Thus,
the field and temperature dependence of the critical current is

\begin{equation}
J_{c}(H,T)\propto H^{-5/6}\Theta \label{jc}
\end{equation}

Most of the experiments reported in the literature have been made
at constant temperature varying field. In this case the theory
predicts that the critical current decreases as $(1-b)$ and the
rapid change of $J_{c}(H,T)$ as a function of $b$ has been
a\-sso\-cia\-ted to this field dependence. However, this behavior
is difficult to check due to the narrow available experimental
field range between $H_{p}(T)$ and $H_{c2}(T)$.

In order to verify the behavior predicted by Eq. \ref{jc} we
investigated both, the field as well as the temperature dependence
of $J_c(H,T)$. It should be noticed that Eq. \ref{jc} has an
explicit dependence in magnetic field when \mbox{$J_{c}(H,T)$} is
plotted as a function of $\Theta $. This dependence should be
evident when measuring the critical current as a function of
temperature for different fields. In this way it can be proved
whether the vortex structure corresponds to an homogeneous vortex
phase described by a correlation volume limited in its transverse
direction, $R_{c}\!=\!a_{0}$.

\section{Experimental}

The experiments were performed in pure and Fe doped NbSe$_{2}$
single crystals. Material parameters for the Fe doped sample are
$T_{c}\,=\,5.8\,$K, $\delta\,T_{c}(10\%\,-\,90\%)\,=\,0.2\,$K,
$\xi(0)\,=\,78\,$\AA and for the pure sample $T_{c}\,=\,7.02\,$K,
$\delta\,T_{c}(10\%\,-\,90\%)\,=\,0.1\,$K, $\xi(0)\,=\,89\,$\AA.

The results for the pure and impure samples are qualitatively
similar. However, measurements of the Fe doped sample have a
relative advantage over the pure one due to the wider field region
$H_{p}(T)\!<\!H\!<\!H_{c2}(T)$.

The transport properties were measured using the Corbino-contact
configuration as shown in the insert of Fig. 6 (b). A dc current
was injected between the central contact and the four contacts
indicated in the figure. The distance between $I ^{-}$ and $V^{-}$
is approximately $0.3\,$mm and between $V^{+}$ and $V^{-}$ is
$0.3\,$mm, in both cases the sample thickness is around $40\,\mu
$m.

The sample was thermally anchored to a sapphire glued to a copper
sample holder that was thermally  weakly connected to a liquid
helium bath at $4.2\,$K. A heater on the sample holder was used to
control the tem\-\-pe\-ra\-tu\-re of the sample by means of a PID
controller.

\begin{figure}[hbb]
\vspace{-1mm}
\includegraphics[width=55mm,angle=-90]{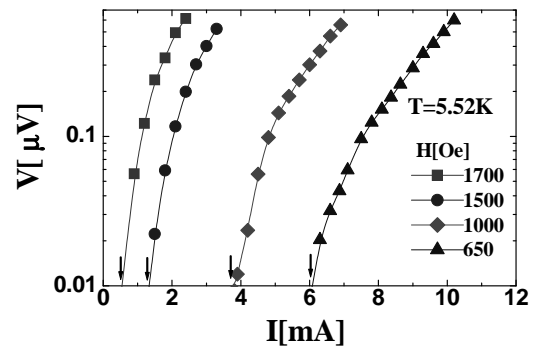}
\vspace{2mm}
 \caption[]{Typical $I-V$ curves for the Fe doped sample
using the Corbino-contact configuration, where the critical
current is marked with an arrow. The measurements were made at
different magnetic fields and $T=5.52\,K$ greater than $T_p(H)$.}
\label{fig:1}
\end{figure}

Measurements were made either cooling the sample in a field, FC
experiments, or cooling the sample in zero field and then sweeping
the field at a given temperature, ZFC.

The critical current was defined as that inducing a vol\-ta\-ge of
$10\,$nV in the Fe doped sample and of $20\,$nV in the pure
sample. Moderate changes in voltage criteria to define
$I_{c}(H,T)$ do not modify the conclusions reached in the paper.

The Bitter decoration in pure samples was performed at $4.2\,$K in
FC experiments following the procedure des\-cri\-bed in Ref.
\onlinecite{marchevsky}. The magnetic decoration of the Fe doped
sample was made at $3\,$K in order to achieve the necessary
resolution. The ZFC decorations of pure samples were made applying
the field after cooling the sample down to $4.2\,$K.

\section{Results and discussion}

Figure \ref{fig:1} shows typical $I\!-\!V$ curves for different
mag\-ne

\begin{figure}[htb]
\includegraphics[width=80mm]{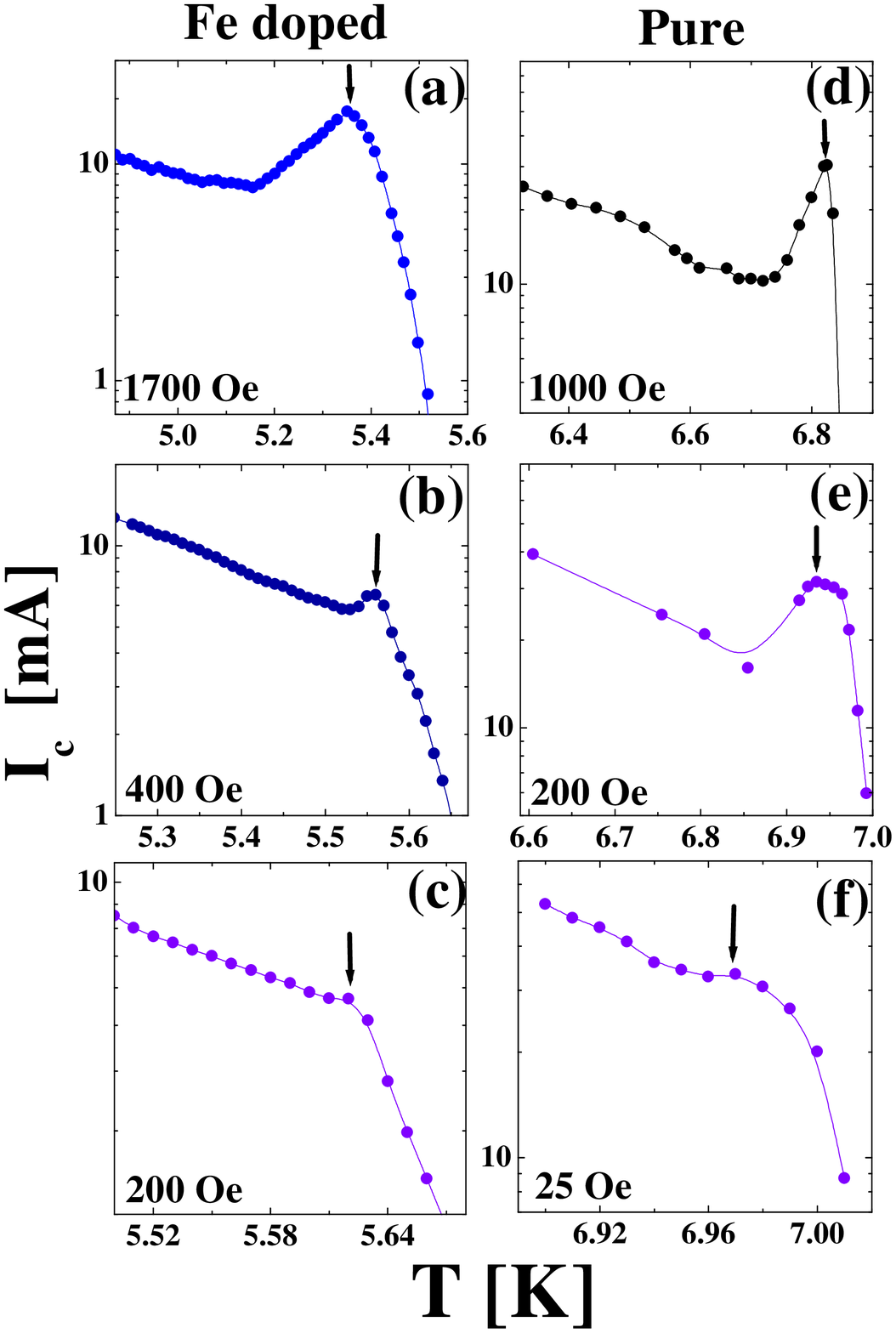}
\vspace{2mm}
 \caption[]{ Critical current curves for different magnetic
fields as a function of temperature for the Fe doped sample: (a),
(b), (c) and for the pure sample: (d), (e), (f). The position of
the peak in $I_{c}(H,T)$ is indicated by an arrow. Figures (c) and
(f) show the $I_{c}(H,T)$ curves at the lowest field where
 the peak is detected for each sample.}
\label{fig:2}
\end{figure}

\noindent \-tic fields obtained in the Corbino contact
configuration for the Fe doped sample in the temperature range
$T\!>\!T_p(H)$. The critical current for each magnetic field is
marked with an arrow. Similar results are obtained for

\noindent the pure sample. For the same value of $I\!-\!I_{c}$ the
voltage increases with $H$, as expected.

Critical current curves as a function of temperature for different
magnetic fields are shown in Fig. \ref{fig:2} for Fe doped and
pure NbSe$_{2}$ samples. A well defined peak in $I_c(H,T)$ is
found for fields higher than $200\,$Oe and $25\,$Oe for the Fe
doped and pure samples respectively (see Fig. \ref{fig:2}(c) and
(f)). Bellow those fields only a change in the slope of
$I_{c}(H,T)$ is detected when approaching $T_{c2}(H)$. We have
defined $H_{c2}(T)$ (or $T_{c2}(H)$) as the magnetic field (or
temperature) where the critical current extrapolates to zero. The
results are shown in Fig. \ref{fig:6} (a) and (b).

\begin{figure}[bbb]
\vspace{-1mm}
\includegraphics[width=60mm,angle=-90]{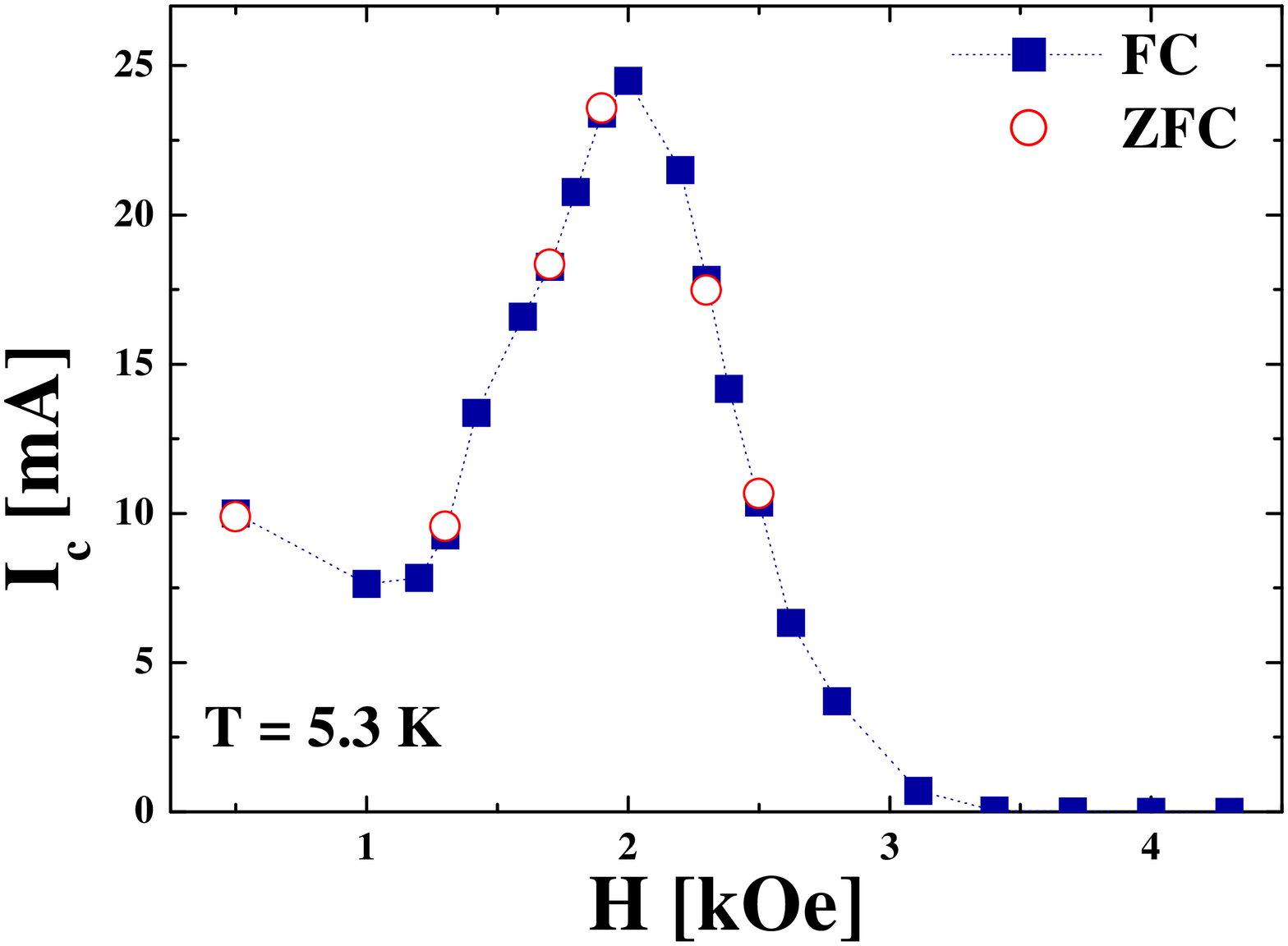}
\vspace{2mm} \caption[]{Field dependence of $I_c(H)$ at $T=5.3\,$K
in the Fe doped sample. The filled square symbols depict the peak
effect for FC measurements while the open circles are for the ZFC
case. } \label{fig:3}
\end{figure}

The critical current of both materials obtained with the described
experimental setup is history independent and consequently
$I_c(H,T)$ is a single-valued function of $H$ and
 $T$. An example is shown in Fig. \ref{fig:3} for the Fe doped sample. Most
of the data reported in this work was taken sweeping temperature.

\begin{figure}[htb]
\vspace{-1mm}
\includegraphics[width=80mm]{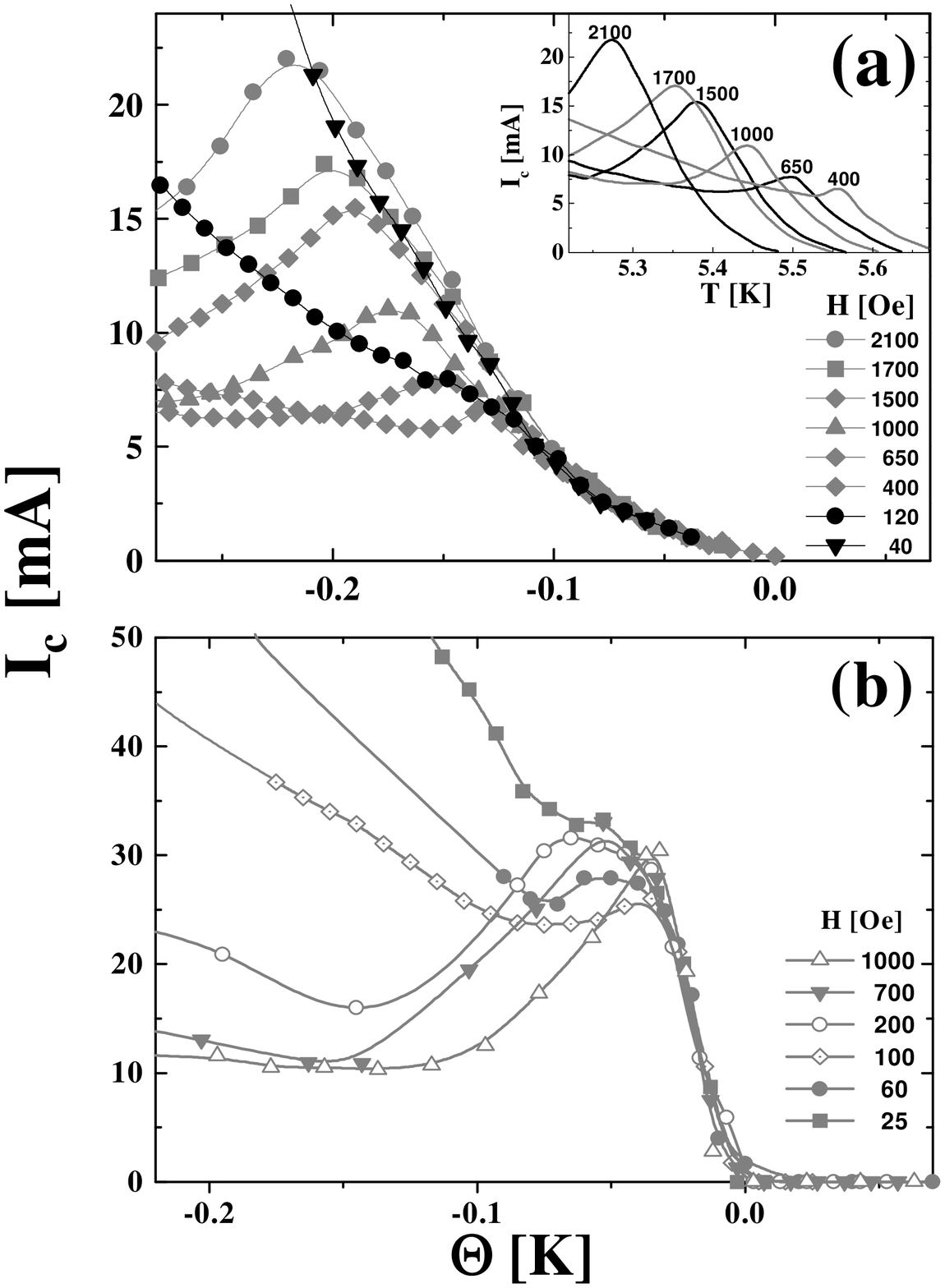}
\caption[]{Critical current as a function of the temperature
$\Theta$ (see text) for different magnetic fields. (a) Fe doped
sample: Dark symbols correspond to the results where no peak in
$I_{c}(H,\Theta)$ is detected. The insert shows the same curves as
a function of the absolute temperature. (b) Pure sample. The data
shows no agreement with the LO theory, see text. } \label{fig:4}
\end{figure}

\begin{figure}[htb]
\vspace{-2mm}
\includegraphics[width=80mm]{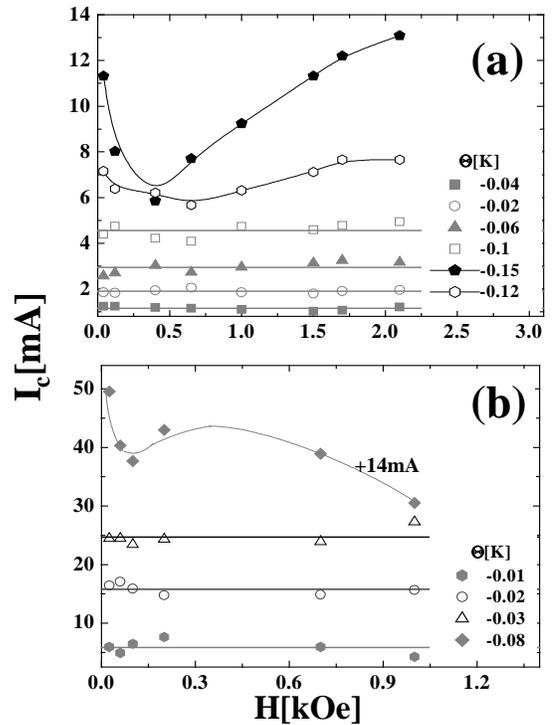}
\vspace{2mm}
 \caption[]{ Plot of $I_{c}(H,\Theta)$ as a function of $H$ for different
$\Theta$. For the Fe doped sample (a) the single pinning regime is
found for values of $\Theta\!\geq\!-0.1$ and for the pure sample
(b) for $\Theta\!\geq\!-0.03$. } \label{fig:5}
\end{figure}

\begin{figure}[htt]
\includegraphics[width=85mm]{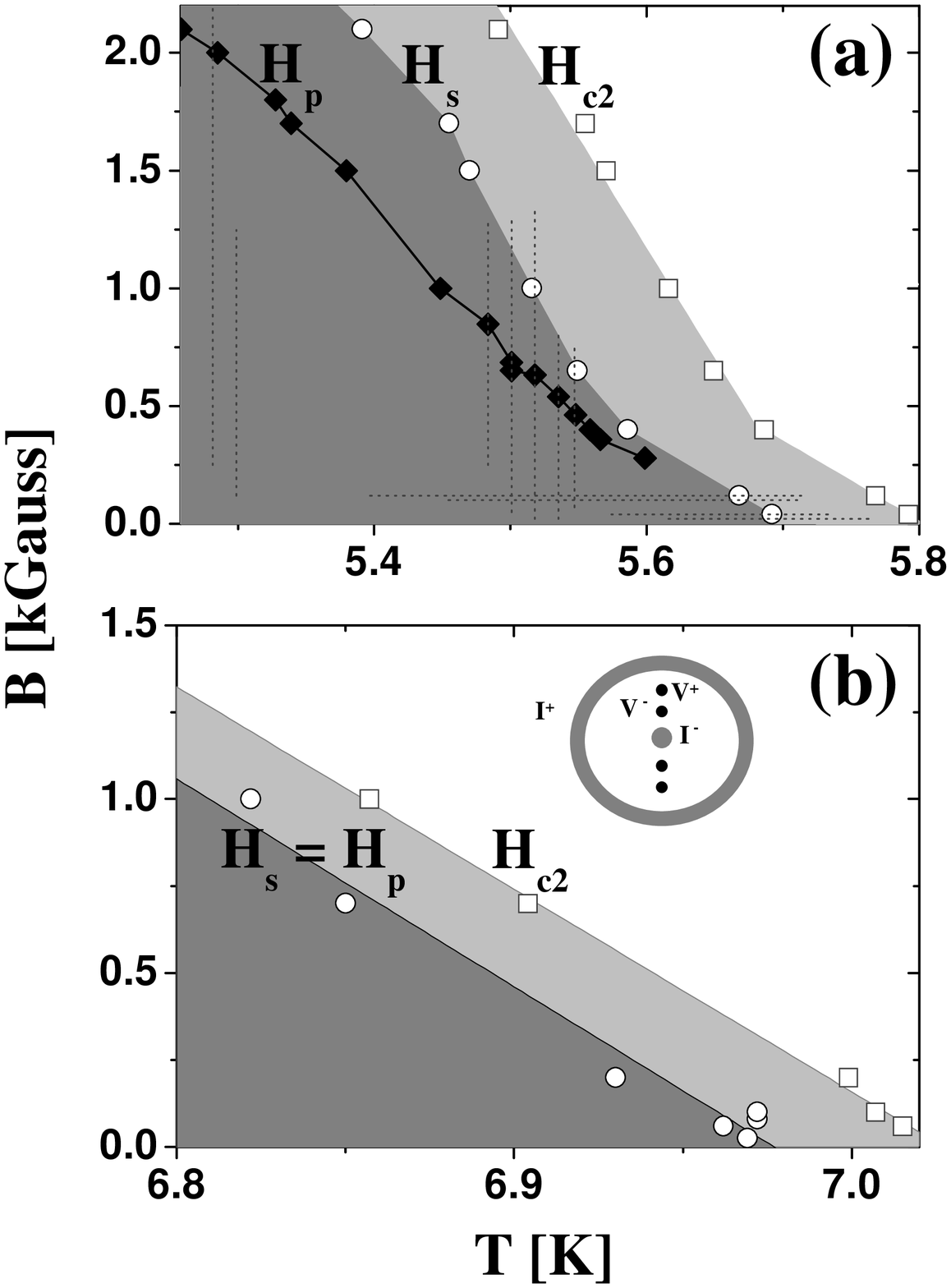}

 \caption[]{  $B\!-\!T$  phase  diagrams. (a)  Fe doped   sample
($T_c\!=\!5.8\,$K). $H_{c2}(T)$ is the field where $I_c$ goes to
zero. The $H_p(T)$ line is the position of the peak of $I_c(T,H)$.
The line $H_s(T)$ corresponds to a crossover to a single pinning
regime. Dotted lines indicate the paths followed  to study the
reentrance of the disordered vortex phase. (b)   Pure sample
($T_c\!=\!7.02\,$K). In this case the $H_s(T)$ line coincides with
$H_{p}(T)$. In the insert is shown the Corbino contact
configuration. } \label{fig:6}
\end{figure}

To verify whether the LO theory (Eq. \ref{jc}) describes the field
and temperature dependence of the results shown in Fig.
\ref{fig:2}, the critical current for different fields is plotted
as a function of the variable $\Theta $, see Fig. \ref{fig:4}.
Contrary to what is expected from Eq. \ref{jc}, the critical
current in the variable $\Theta$ becomes field  independent in a
region of temperatures higher than that where the peak in
I$_c(H,T)$ takes place : $-0.1\leq\Theta\leq 0$ for the Fe doped
sample and $-0.03\leq\Theta\leq 0$ for the pure case. This is
clearly shown in Fig. \ref{fig:5} where $I_{c}(H, \Theta)$ as a
function of magnetic field for different values of $\Theta$ is
plotted. This field independent regime for the critical current is
obeyed even for the range of fields where the peak effect is not
observed, see Fig. \ref{fig:4}(a).

As a consequence of the result described in the pre\-vious
paragraph it is concluded that the LO theory (see Eq.\ref{jc}) is
not applicable in the $\Theta$ region where the cri\-ti\-cal
current is field independent. Besides this, the field dependence
of $I_{c}(H,\Theta)$ at the peak does not follow the behavior
predicted by the theory in the limit $R_{c}\!=\!a_{0}$. The
predicted increase of $I_{c}(H,\Theta)$ at the peak when
de\-crea\-sing the magnetic field is  not observed in any of the
samples. Thus, the behavior of the critical current in the peak
effect region is not qualitatively described by the LO collective
theory under the assumption $R_{c}\!=\!a_{0}$.

The results in Figs. \ref{fig:4} and \ref{fig:5} make evident that
the field and temperature dependence of the critical current
 are only determined by the change of superfluid density as described by $\lambda
_{h}^{-2}(T,H)\,\propto\,\Psi^{2}\,\propto\,(1-t)(1-b)\!=\!\Theta\,T_{c}$,
when approaching $T_{c2}(H)$. The lack of explicit field
dependence of $I_{c}(H,\Theta)$ indicates the irrelevance of the
vortex-vortex interaction to determine the pinning energy,
su\-gges\-ting that the critical current is fully determined by
the pinning of individual vortices: single vortex limit. This
seems to be a counter-intuitive and surprising re\-sult in a field
region where the distance between cores is smaller than the
interaction cha\-rac\-te\-ris\-tic length,

\noindent $\lambda _{h}(T,H)$. We will come back to this point in
the conclu\-sions.

Figure \ref{fig:6} summarizes the results described up to now in a
$B\!-\!T$ phase diagram. In Fig. \ref{fig:6} (a) the line
$H_{p}(T)$ indi\-ca\-tes the field where the peak in $I_{c}(H,T)$
takes place and the line $H_{s}(T)$ marks the crossover to the
single vor\-tex regime (lack of field dependence of
$I_{c}(H,\Theta)$)  for the Fe doped sample. The Fig. \ref{fig:6}
(b) shows the results for the pure sample, in this case $H_{s}(T)$
coincides with $H_{p}(T)$. The new individual vortex regime line
$H_{s}(T)$ is found to follow the condition
$\Theta_{s}\!=\!T_{s}(H)-T_{c2}(H)\!=\!constant$.

In order to detect a possible reentrance of the line $H_{p}(T)$ in
the phase diagram of the Fe doped sample as reported in Ref.
\onlinecite{corbino}, we have made careful measurements of
critical current as a function of temperature or magnetic field
keeping constant field or temperature respectively. No evidence of
a reentrant $H_{p}(T)$ was found for the  investigated sample (the
paths followed to detect the possible reentrance are indicated
with dotted lines in Fig. \ref{fig:6}(a)). The presence of a peak
effect down to $200\,$Oe and the absence of reentrance in the Fe
doped sample is a di\-ffe\-rent behavior than that found for
similar samples in Ref. \onlinecite{corbino} (comparable $T_{c}$
and $\xi(0)$) where the reentrance of the high temperature phase
takes place at $H\!=\!1000\,$Oe.

\noindent This clearly shows that the reentrant peak effect
depends strongly on unknown properties of the sample and
conse\-quently on the ubiquity of the supposed disordered phase as
has been verified by several authors.\cite{discusion}

\begin{figure}[htb]
\vspace{2mm}
\includegraphics[width=75mm]{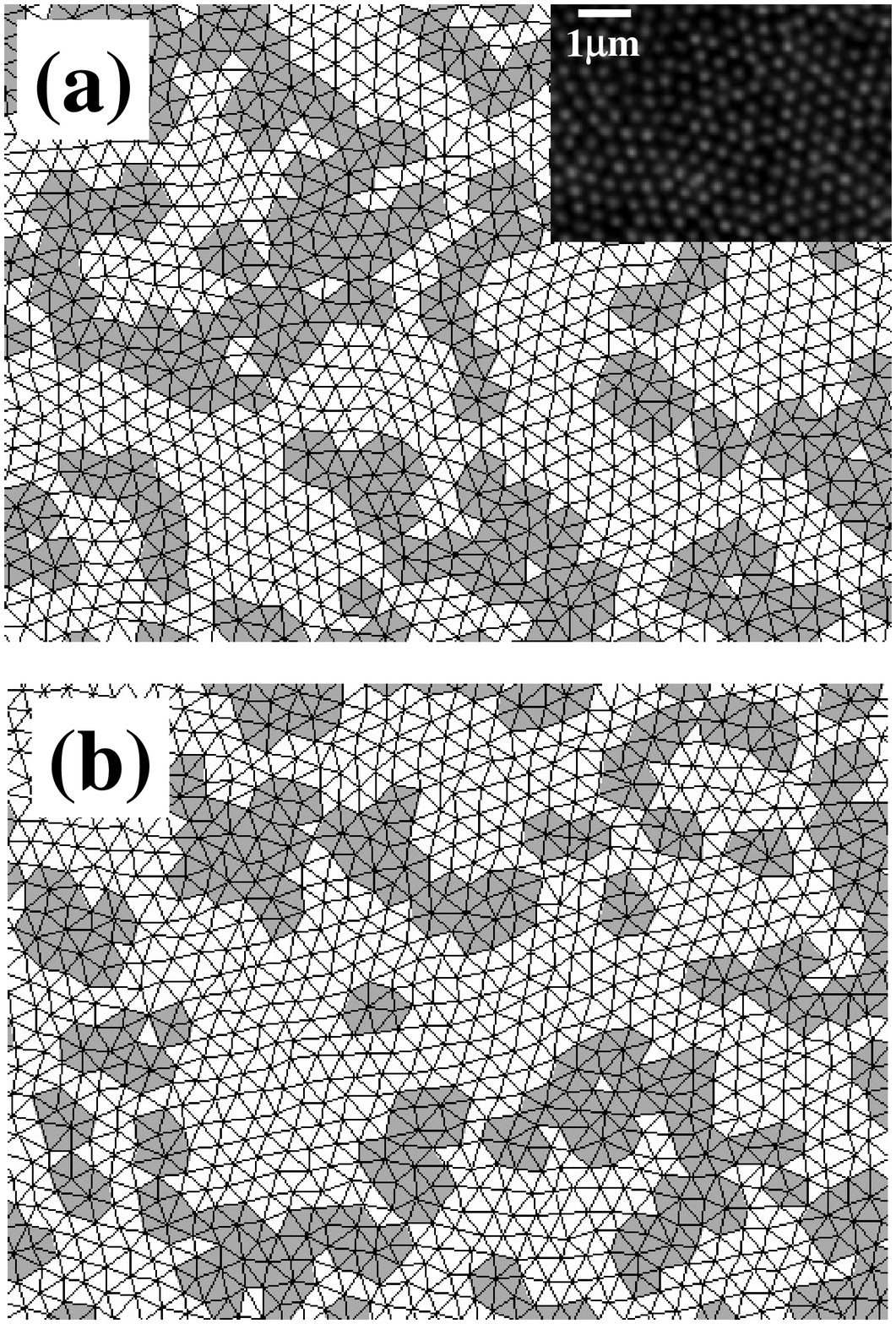}
\vspace{7mm} \caption[]{FC vortex structure in NbSe$_{2}$ observed
by magnetic decoration at $60\,$Oe. Delaunay triangulations where
the nearest neighbors vortices are bonded and the non-sixfold
coordinated are marked in gray. (a) Pure sample decorated at
$4.2\,$K and (b) Fe doped sample decorated at $3\,$K. }
\label{fig:7}
\end{figure}

The results of magnetic decorations performed in FC experiments at
$60\,$Oe  are shown in Fig. \ref{fig:7}(a) for the pure sample at
$4.2\,$K and in (b) for Fe doped sample at $3\,$K. It is important
to remark that in this case the FC decoration for the Fe doped
sample is made at a field below the lowest field where the peak is
detected. In the pure case the FC at $60\,$Oe crosses the $H_s(T)$
line. An identical polycrystalline nature of the FC state is
observed in both type of materials. Following
Ref.\onlinecite{xiao},  this structure co\-rres\-ponds to the so
called metastable disordered phase. The polycrystalline structure
is detected  independently of the presence of the peak effect at
the field where decoration is made. This indicates that the FC
polycrystalline structure is originated in the region of
temperatures where  the single vortex regime dominates. This is an
interesting result showing that some degree

\noindent of orientational order is preserved when nucleating the
vortex structure at $H_{c2}(T)$ in a single vortex limit.

The Bitter structure obtained in ZFC experiments at $4.2\,$K shows
gradients of $B$ as a result of the presence of transport currents
induced when applying the field.\cite{marchevsky} Despite this, a
polycristalline structure is found in most of the sample. The
average size of the grains coincides with those shown in Fig.
\ref{fig:7} obtained in FC experiments.

\section{Conclusions}

We have found a universal behavior of the temperature and field
dependent critical current when a\-pproa\-ching the mean field
superconducting transition, $T_{c2}(H)$. The universal character
is made evident when plotting $I_{c}(H,T)$ as a function of the
temperature difference $\Theta\!=\!T\,-\,T_{c2}(H)$, as shown in
Figs. \ref{fig:4} and \ref{fig:5}. This shows that the superfluid
density univocally determines the pinning properties of the
material in this region of the phase diagram. The lack of explicit
field dependence of $I_{c}(H,\Theta)$ when plotted as a function
of $\Theta$ indicates that the vortex-vortex interaction plays no
significant role in the determination of vortex pinning. The
results cannot be interpreted within the collective pinning theory
in the single vortex limit\cite{LO} since the distance between
vortex cores is much smaller than the effective penetration depth,
$\lambda_{h}$. This rather suggests that in this large
$\lambda_{h}$ limit, the interaction of individual pinning centers
with a vortex line overcomes the weak interaction between
vortices.\cite{brandt}

It is important to remark that the novel single pinning regime is
not necessarily associated with the presence of the peak effect:
despite the peak effect is not detected below $200\,$Oe in the Fe
doped sample the crossover to single pinning takes place even at
lower fields.

The transport and Bitter experiments make evident that the single
vortex pinning found at high temperatures is compatible with a
structure where directional order is preserved within the grains.
This reinforces the claim that the typical length scale associated
with critical current is different from that characterizing the
topological order.\cite{giamarchi}

We have not been able to detect a reentrance of $H_{p}(T)$,
neither in the Fe doped sample nor in the pure one. Whether this
indicates that the phenomenon fades away when decreasing field or
transforms into a weak signal not detected in our samples deserves
further research.

The combined transport and Bitter decoration experiments together
with the results in Ref.\onlinecite{xiao}  open a question mark on
the nature of the disordered phase. This is of particular
importance in view of recent data
interpreted\cite{corbino,vinokur,brandt2001} in terms of an
order-disorder thermodynamic transition that takes place at the
temperature where the peak in $I_c(H,T)$ is observed. The results
reported in this paper call for a better definition of what is
meant by order and disorder characterized by critical current
measurements. The topology of the FC structure as observed by
Bitter decoration shows no gradient of $B$ indicating the absence
of transport currents acting on the bulk of the vortex
system.\cite{victor,unpublished} Although this supports that by FC
the high temperature structure is supercooled, the results make
evident that the high temperature phase is not amorphous.

The observed ZFC vortex structure is not an ordered equilibrium
state; it is consistent with a topological unstable state
dominated by the presence of transport currents. \cite{marchevsky}

In conclusion, through the analysis of the vortex structure and
the assumption of the existence of the proposed supercooling of
the high temperature phase in the FC experiments, no evidence of
an order-disorder phase transition is found in samples where the
transport properties show anomalies associated with a first order
thermodynamic transition.\cite{corbino}

\section{Acknowledgments}

We would like to acknowledge D. L\'opez and H. Safar for
stimulating discussions and P. Gammel and D. J. Bishop for
providing the samples. We would also like to thank P. Pedrazzini
for a careful reading of the manuscript.

Work partially supported by ANPCYT PICT99-5117, CONICET
PIP96/4207, Fundaci\'on Balseiro and Fundaci\'on Antorchas
1370/1-1118. M. M. thanks CONICET and Y. F. thanks ANPCYT for financial support.


\begin{references}
\bibitem{beqh}In the range of fields where the experiments are performed
and due to the geometrical aspects of the samples, it is a good
approximation to take the induction field, $B$, equal to the
applied field, $H$. Moreover, the magnetic decoration technique
shows that the difference between the a\-pplied field and the
magnetic induction is always less than $10\%$.
\end{references}
\end{document}